\documentclass[aps,prl,twocolumn,groupedaddress,showpacs]{revtex4}

\usepackage{graphicx}
\usepackage{dcolumn}
\usepackage{bm}

\begin{document}


\title{Deformation of a Trapped Fermi Gas with Unequal Spin Populations}



\author{G. B. Partridge}
\author{Wenhui Li}%
\author{Y. A. Liao}
\author{R. G. Hulet}
\affiliation{Department of Physics and Astronomy and Rice Quantum
Institute, Rice University, Houston, TX 77251, USA.
}%

\author{M. Haque}
\author{H. T. C. Stoof}
\affiliation{Institute for Theoretical Physics, Utrecht University, Leuvenlaan 4, 3584 CE Utrecht, The Netherlands.}%

\affiliation{}


\date{\today}

\begin{abstract}
The real-space densities of a polarized strongly-interacting two-component Fermi gas of $^6$Li atoms reveal two low temperature regimes, both with
a fully-paired core. At the lowest temperatures, the unpolarized core deforms with increasing polarization. Sharp boundaries between the core and
the excess unpaired atoms are consistent with a phase separation driven by a first-order phase transition. In contrast, at higher temperatures the
core does not deform but remains unpolarized up to a critical polarization. The boundaries are not sharp in this case, indicating a
partially-polarized shell between the core and the unpaired atoms.  The temperature dependence is consistent with a tricritical point in the phase
diagram.
\end{abstract}

\pacs{03.75.Ss, 05.70.Fh, 74.25.Dw}

\maketitle
The formation of pairs consisting of one spin-up and one spin-down electron underlies the phenomenon of superconductivity. While the populations
of the two spin components are generally equal in superconductors, an imbalance is readily produced in experiments with gases of trapped,
ultracold fermionic atoms, as was recently demonstrated \cite{Zwierlein06,Partridge06}. Exotic new states of matter are predicted for the
unbalanced system that, if realized, may have important implications for our understanding of nuclei, compact stars, and quantum chromodynamics.
Calculations show that phase separation between pairs and the excess unpaired atoms is one possible outcome for a strongly interacting
two-component gas \cite{Bedaque03, Carlson05, Sheehy06,Gu06, Hu06}.  We previously reported evidence for a phase separation in a trapped atomic
Fermi gas to a state containing a paired central core, with the excess unpaired atoms residing outside this core \cite{Partridge06}.  Such a phase
separation can be detected in the real-space distributions using \emph{in-situ} imaging, where a uniformly paired region produces a minimum in the
difference distribution obtained by subtracting the majority and minority spin densities \cite{Partridge06}.

In our previous work, we noted that the excess unpaired atoms reside primarily at the axial poles of the highly-elongated trap, while relatively
few occupy the equatorial shell.  As a result, the central minimum in the difference images was accompanied by a corresponding central dip in the
axial density profile obtained by integrating the two-dimensional column density along the radial coordinate \cite{Partridge06}. Several authors
have calculated spatial distributions for phase separation, assuming both a harmonic trapping potential and the local density approximation (LDA)
\cite{Pieri06, Yi06, Chevy06, Pao06, Chien06}. It was pointed out that, under these assumptions, a uniformly paired core would produce a constant
axial density difference, rather than a central dip \cite{DeSilva06,Haque06}. Shin \emph{et al.} have recently adopted \emph{in-situ} imaging, and
present their resulting images as evidence for phase separation, though in their experiment, no such deformation is observed \cite{Shin06}.

In this paper, we characterize the properties of the phase-separated state using detailed quantitative measurements of the deformation and
reconstructions of the three-dimensional (3D) density distributions. Furthermore, we explore the role that finite temperature plays in determining
the overall behavior of this system and identify two distinct low-temperature regimes.

Our methods for producing a strongly-interacting, two-component Fermi gas of $^6$Li atoms have been described previously
\cite{Partridge05,Partridge06}.  The relative population of two hyperfine states, designated as $|1\rangle$ and $|2\rangle$, is controlled by
driving radio-frequency transitions between them. Spin relaxation is negligible over the duration of the experiment. A nearly uniform magnetic
field is tuned to the location of a broad Feshbach resonance at 834 G \cite{Houbiers98, Bartenstein05}, where the two-body scattering length
diverges ($\pm \infty$) producing unitarity-limited strong interactions.  The combined optical and magnetic trapping potential is given by $
U(r,z)
=\frac{1}{2}m\omega_B^2z^2 + U_o\big(1-\big(\frac{w_o^2}{w^2(z)}\big) e^{-2r^2/w^2(z)}\big)$
, where $\omega_B=(2\pi)$ 6.5 Hz, $w_o$ = 26 $\mu$m, $w(z) = w_o[1 - (z/z_o)^2]$, and $z_o$ = 1.7 mm. Both radial and axial potentials are
approximately harmonic for sufficiently small $r$ and $z$.  The atoms are evaporatively cooled by reducing the trap laser intensity until $U_o$
achieves its final value of 540 nK.  At this trap depth, the radial and axial trap frequencies are $\omega_r=(2\pi)$ 325 Hz and $\omega_z=(2\pi)$
7.2 Hz, respectively. The two states, $|1\rangle$ and $|2\rangle$, are sequentially imaged in the trap by absorption. The first optical probe
pulse breaks pairs, causing a small heating that radially broadens the second images. We have reduced the delay between probe pulses to 27 $\mu$s,
significantly reducing the probe-induced broadening previously observed for delays of 215 $\mu$s \cite{Partridge06}. Analysis of the images
provides measurement of the number of atoms in each state, $N_1$ and $N_2$, from which the polarization $P = (N_1 - N_2)/(N_1 + N_2)$ is
determined. Fitting the profiles of gases deliberately prepared as $P = 0$ to fermionic nonzero-temperature Thomas-Fermi distributions, gives
fitted temperatures of $\widetilde{T} \lesssim 0.05$ $T_F$, where $T_F$ is the Fermi temperature. The actual temperatures are expected to be
closely related to $\widetilde{T} $ \cite{Kinast05}.
 \begin{figure}
\includegraphics[scale=1.0, bb =190 256 418 532]{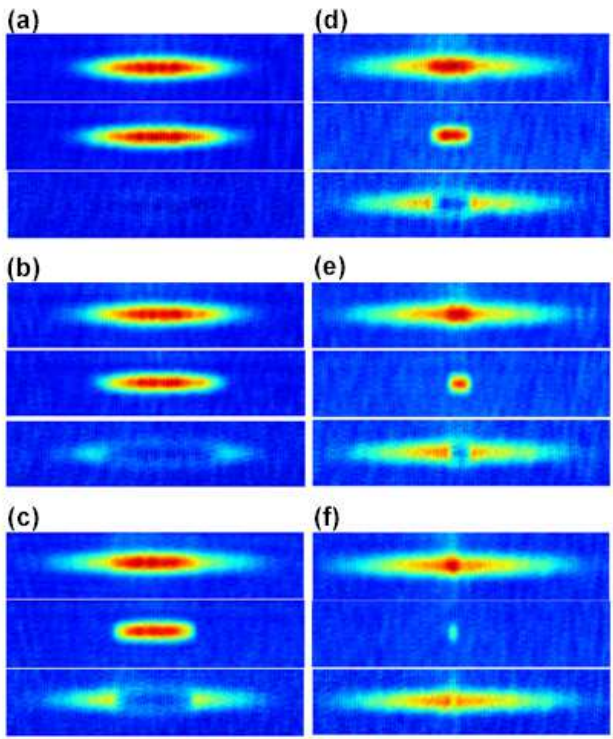}
\caption{\label{fig:fig1_pics} \textit{In-situ} absorption images of a polarized Fermi gas.  The top figure of each sequence corresponds to the
column density of state $|1\rangle$, the middle to state $|2\rangle$, and the third to the difference of the two. The polarizations are (a) $P =
0$, (b) $P = 0.18$, (c) $P = 0.37$, (d) $P = 0.60$, (e) $P = 0.79$, and (f) $P = 0.95$. In each sequence, state $|2\rangle$ was imaged first,
followed by state $|1\rangle$. Slight probe-induced heating can be discerned in the images of state $|1\rangle$, where the distribution bulges
slightly in the radial direction in the region of overlap between the two states. The field of view for these images is 1654 $\mu$m by 81 $\mu$m.
The displayed aspect ratio was reduced by a factor of 4.4 for clarity. }
\end{figure}
 \begin{figure}
\includegraphics[scale=1.0, bb= 17 17 262 216]{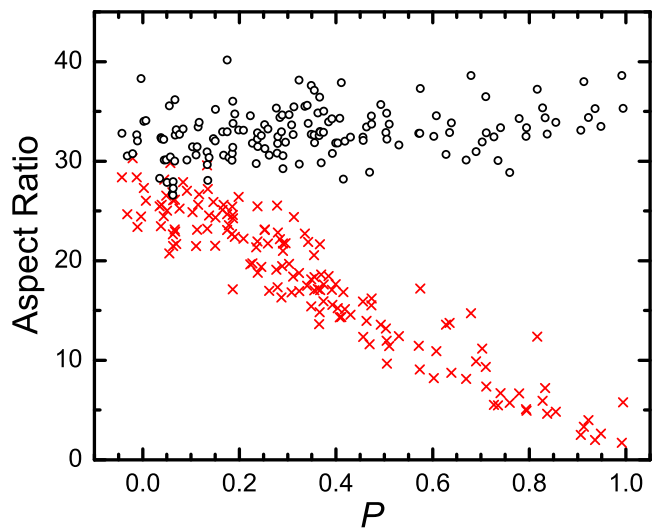}
\caption{\label{fig:fig2_ar} Aspect ratio vs. \emph{P}. The ratio of the axial to the radial dimensions, $R_z/R_r$, is shown for state $|1\rangle$
by the black circles and for state $|2\rangle$ by the red crosses.  The radii $R_r$ for both states are determined by fitting the column density
profiles to zero-temperature, fermionic Thomas-Fermi distributions.  The axial distributions are distinctly non-Thomas-Fermi-like, so $R_z$ is
found by a simple linear extrapolation of the column density to zero.  An aspect ratio of 36 is the expected value for a non-interacting gas with
anharmonic corrections, in reasonable agreement with the observations.  The uncertainty in $P$ is 0.04, which is the standard deviation of
polarization measurements deliberately prepared as $P = 0$.  There are shot-to-shot variations in $N_1$ and a small systematic variation towards
larger $N_1$ at smaller $P$.  For $P < 0.40$, $N_1$ = 170 k $\pm$ 40 k, and for $P > 0.40$, $N_1$ = 135 k $\pm$ 25 k, where the uncertainty is the
standard deviation of the measurements.  The corresponding average Fermi temperature is $T_F \approx 430$ nK, where we define $T_F =
\hbar(\omega_r^2 \omega_z)^{1/3}(6 N_1)^{1/3}/k_B$. }
\end{figure}

Figure 1 shows a series of images corresponding to a range of $P$ from 0 to 0.95.  The minority spin ($|2\rangle$) distribution becomes markedly
less elongated with increasing $P$, while its radial size remains approximately the same as that for the majority spin ($|1\rangle$). This
deformation causes the bunching of unpaired atoms at the axial poles, and a lack of them in the equatorial shell. Remarkably, the deformation
grows continuously with \emph{P}.  The central holes in the difference distributions, which are approximately equal to the background level for
all but the highest values of $P$, indicate that the central core is nearly uniformly paired. Figure 2 shows the aspect ratio for both states.
While the majority state aspect ratio changes little, that of the minority (representing the core) decreases by a factor of 10 when going from
completely unpolarized ($P = 0$) to completely polarized ($P = 1$).

Figure 3(a) presents a center-line cut of the column densities of the majority and minority states, as well as their difference, taken along the
axial direction.  The axial density difference profile of this data also exhibits a pronounced central dip, as in Ref. \cite{Partridge06}.  A
sharp phase boundary between the core and excess fermions is readily apparent in Fig. 3(a), indicating that a partially polarized shell, if it
exists, is extremely thin.  Cylindrical symmetry of the trap enables reconstruction of the true 3D density distribution $n(r, z)$ from the column
densities by use of the inverse Abel transform \cite{Smith88}.  An axial cut of the reconstructed 3D density, $n(r=0, z)$, is shown in Fig. 3(b).
The ratio of the central densities, $n_1(0,0)/n_2(0,0)$, obtained from the reconstructed 3D distributions, are plotted vs. $P$ in Fig. 4(a). The
central core remains unpolarized until at least $P \approx 0.9$, in contradiction to the results reported in Ref. \cite{Shin06}, where uniform
pairing was observed to break down at $P \approx 0.77$.
\begin{figure}
\includegraphics[scale=1.0, bb= 15 15 238 199]{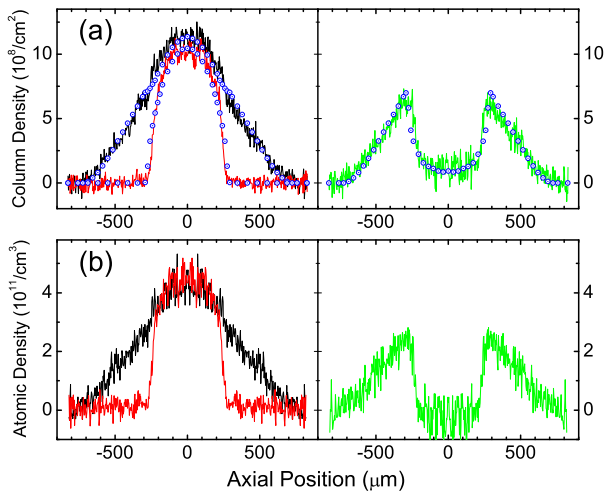}
\caption{\label{fig:fig3_profs} Column-density profile and 3D density reconstruction.  The black lines correspond to state $|1\rangle$, the red to
state $|2\rangle$, and the green to their difference, for $P = 0.35$ and $N_1$ = 175 k.  The circles are the result of our theory (see text). (a)
Center-line ($r = 0$) axial cut of the column-densities. (b) Center-line axial cut of the reconstructed 3D densities.  The signal to noise of (b)
was improved by reflecting and averaging the column density images about both the $r = 0$ and the $z = 0$ planes before reconstruction. The
densities in (b) are slightly diminished due to probe induced radial broadening.}
\end{figure}

The observed LDA-violating deformations are surprising because the radial size of the distributions is about 10 times larger than the inverse
Fermi momentum $k_F$, the expected magnitude of the correlation length. Nonetheless, several LDA-violating mechanisms, enhanced by confinement in
a high aspect ratio trapping potential, may explain these observations. Gradient terms in the Gross-Pitaevskii equation can lead to LDA-violating
deformations on the BEC side of resonance, but the magnitude of the calculated effect is much smaller than we observe at unitarity
\cite{Imambekov06}. De Silva and Mueller have shown that surface tension between the normal and superfluid phases can result in deformations of
the minority component that are quite similar to those observed here \cite{DeSilva06B}. To quantify this, we have developed a theory for the
density profiles in the gas that allows for a deformed local chemical potential that is consistent with the macroscopic deformation of the
minority density profiles.  The results of this calculation, the details of which will be published elsewhere, is shown in Fig. 3(a) and compares
favorably with experiment. Within the limitations of our calculation, we find no evidence for a deformed Fermi surface \cite{Sedrakian05} in the
experimental data.

We had previously found that phase separation occurred only for $P > P_c$, where $P_c \approx 0.1$ \cite{Partridge06}.  For $P < P_c$, the
observations were consistent with a non-phase-separated polarized superfluid. The present data, however, exhibits phase separation for arbitrarily
small $P$.  Since the previous work, we have improved the efficiency of the evaporation trajectory, and now obtain fitted temperatures that are
about half of those previously attained.  This temperature-dependent behavior is consistent with a phase boundary between a phase-separated regime
and a polarized superfluid (Sarma or breached-pair phase) \cite{Sarma63, Houbiers97, Liu03} at nonzero temperature. Such a phase boundary has
recently been discussed in the context of a tricritical point in the phase diagram \cite{Combescot04, Chevy06, Parish06, Gubbels06}.  To test this
hypothesis, we deliberately produced higher temperatures by stopping the evaporation trajectory at a higher trap depth (1.2 $\mu$K), resulting in
$\widetilde{T} \approx 0.2$ $T_F$. Figure 4(b) shows that the central densities in this case remain equal until a critical polarization of $P
\approx 0.6 - 0.7$ is reached.
\begin{figure}
\includegraphics[scale=1.0, bb=13 20 214 216]{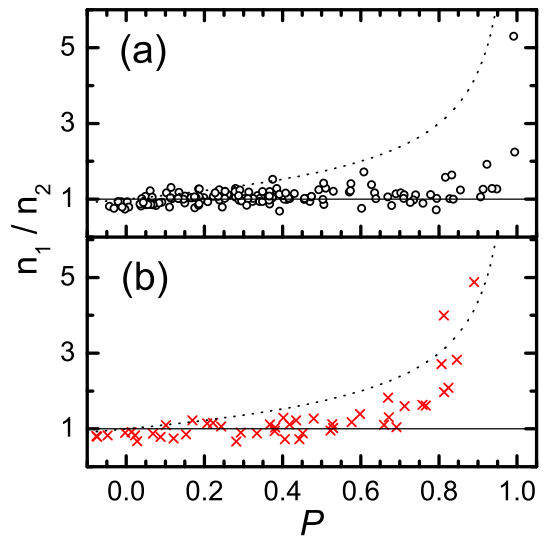}
\caption{\label{fig:fig4_nvp}Ratio of the central densities vs. polarization.   (a) $\widetilde{T} \lesssim 0.05$ $T_F$, corresponding to the data
shown in Fig. 2; (b) $\widetilde{T} \approx 0.2$ $T_F$, with average $N_1$ = 500 k.  The dotted lines correspond to $[(1+P)/(1-P)]^{1/2}$, the
expected central density ratio for a harmonically confined, non-interacting gas at $T = 0$.  The increase in $n_1(0,0)/n_2(0,0)$ in (a) for $P >
0.9$ may be explained by higher temperatures for these data that arise from inefficiencies in evaporative cooling at very high $P$. }
\end{figure}

Figure 5 shows absorption images prepared at both the lower and higher temperatures.  It is readily apparent from the images that the density
distributions of the two components of the higher temperature gas show no deformations, in contrast to those of the colder case.  We find that at
higher temperatures, the aspect ratios of the minority and majority components remain equal and constant for all $P$.  This lack of deformation is
also evident in the axial density distributions, where in the case of the colder data, the axial difference distribution (Fig. 5(b)) shows the
characteristic double-peaked structure observed previously \cite{Partridge06}, while that of the warmer cloud (Fig. 5(d)) exhibits the flat-topped
distribution reported in Ref. \cite{Shin06} and expected for a paired core with no deformation \cite{DeSilva06, Haque06}. The phase boundary is
also much sharper for the low temperature data. In summary, the higher temperature data support the suggestion of a temperature dependent
transition between a low-temperature phase separated state and a higher temperature polarized superfluid \cite{Parish06, Gubbels06}.
 \begin{figure}
\includegraphics[scale=.65, bb= 125 221 473 576]{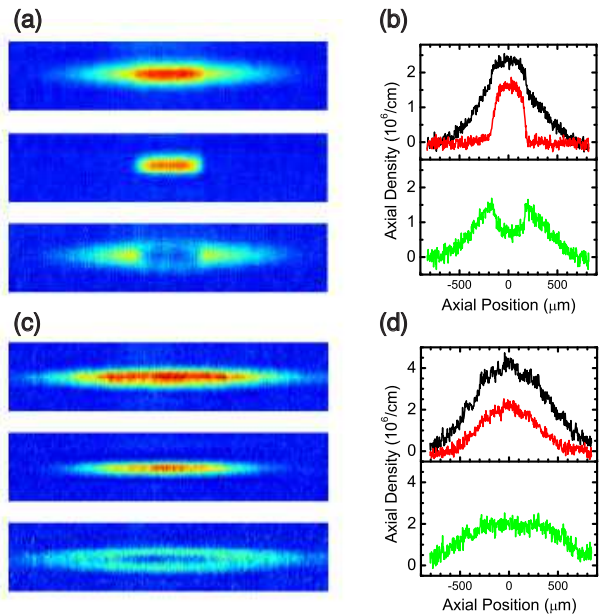}
\caption{\label{fig:fig5_dt}\textit{In-situ} absorption images and integrated profiles.  To the left are absorption images, while the plots to the
right are the corresponding axial density distributions.  (a),(b): $P = 0.50$, $N_1$ = 146 k, with $\widetilde{T} \lesssim 0.05$ $T_F$; (c),(d):
$P = 0.45$, $N_1$ = 374 k, with $\widetilde{T} \approx 0.2$ $T_F$. The central bulge evident in the state $|1\rangle$ density in (b) is indicative
of strong attraction between atoms in the core region.}
\end{figure}

We have reported that pairing with unequal spin populations leads to real-space deformations in a highly elongated, but still three-dimensional
geometry.  The sharp phase boundaries between the superfluid core and the polarized normal phase are consistent with the usual convention that
phase separation is associated with first-order phase transitions.  At elevated temperatures, but still below the transition to the normal state,
deformations are absent, and a partially polarized shell is observed between a uniformly paired core and the fully polarized outer shell. These
observations support the existence of a tri-critical point in the phase diagram where the superfluid-normal transition changes from second-order
to first-order as the temperature is lowered.
\begin{acknowledgments}
We thank E. Mueller for helpful discussions and for providing the Abel transform code.  Support was provided by the NSF, NASA, ONR, the Welch
Foundation, the Stichting voor Fundamenteel Onderzoek der Materie (FOM), and the Nederlandse Organisatie voor Wetenschappelijk Onderzoek (NWO).
\end{acknowledgments}


\end{document}